\RequirePackage{fix-cm}
\documentclass[twocolumn,epjc3]{svjour3}  
\smartqed  

\usepackage[utf8]{inputenc}
\usepackage{graphicx}
\usepackage{url}
\usepackage[numbers]{natbib} 
\usepackage{amssymb}
\usepackage{float}
\usepackage{placeins}
\RequirePackage{graphicx}
\usepackage{graphicx}
\usepackage{subcaption} 
\usepackage{amsmath}
\usepackage{caption}    
\usepackage[colorlinks,citecolor=blue,urlcolor=blue,linkcolor=blue]{hyperref}
\usepackage{microtype}
\usepackage{booktabs}
\usepackage{multirow}
\DeclareUnicodeCharacter{202F}{\,}
\journalname{Eur. Phys. J. C}

\begin{document}

\title{Data-driven modeling of Galactic diffuse emission with multi-wavelength observations}  

\author{%
  Xi Liu,\thanksref{addr1}
  Xiaodong Li,\thanksref{e1,addr1}
  Sujie Lin,\thanksref{e2,addr1}
  Yihan Liu,\thanksref{addr1}
  Chengyu Shao,\thanksref{addr1}
  Lili Yang,\thanksref{e3,addr1, addr2}
  Le Zhang\thanksref{addr1}
  }%

\thankstext{e1}{e-mail: lixiaod25@mail.sysu.edu.cn}
\thankstext{e2}{e-mail: linsj6@mail.sysu.edu.cn}
\thankstext{e3}{e-mail: yanglli5@mail.sysu.edu.cn}

\institute{
School of Physics and Astronomy, Sun Yat-sen University, Zhuhai, Guangdong, 519082, China
\label{addr1}
\and
\text{ }Department of Physics, University of Johannesburg, PO Box 524, Auckland Park 2006, South Africa 
\label{addr2}
}

\date{Received: date / Accepted: date}

\maketitle

\begin{abstract}
We present a data-driven investigation of Galactic diffuse emission. Using multi-frequency Planck radio/microwave maps (30–857 GHz) and \textit{Fermi}-LAT gamma-ray data (50 MeV–814 GeV), we construct a nonlinear mapping between radio emission and gamma-ray intensity through supervised machine learning.
Our models achieve high predictive accuracy ($R^2 >$ 0.90 in the 0.1–10 GeV range), demonstrating that multi-frequency radio observations encode sufficient information to reconstruct both spatial morphology and spectral properties of diffuse gamma-ray emission.
By analyzing model performance across different frequency bands and spatial regions, we identify high-frequency radio bands as the dominant predictor, providing direct empirical support for the hadronic origin of Galactic 0.1 -- 10 GeV gamma rays, while low-frequency radio bands for the leptonic origin above 10 GeV.
Residual maps reveal coherent large-scale structures, including Loop I and III, highlighting regions where standard interstellar emission models are incomplete or biased. Compared with the GALPROP model, our machine learning approach yields a higher $R^2=0.95$ and lower mean absolute relative error (14.7\%) in the inner Galactic disk and the Galactic center region. Our results illustrate that machine learning serves as a physically interpretable tool for multi-messenger astrophysics, providing a data-driven baseline for separating non-standard emission components and deriving new constraints on cosmic-ray propagation and interstellar medium structure.

\end{abstract}
\keywords{Galactic diffuse emission \and Machine learning methods \and Multi-wavelength astronomy \and Gamma rays \and Radio continuum}

\section{Introduction} \label{intro}
Understanding Galactic diffuse emission (GDE) is essential for probing cosmic-ray propagation, interstellar medium (ISM) structure, and high-energy processes in our Milky Way. Multi-messenger observations, including cosmic rays, photons, and neutrinos, are intrinsically linked through the interactions of cosmic rays with gas, magnetic, and radiation fields \cite{Planck2015-X,Strong2007, Acero2016}. However, disentangling these coupled processes remains a major challenge in the field.

Traditional approaches rely on physically motivated models of cosmic-ray transport, which incorporate gas distributions, radiation fields, and magnetic configurations \cite{Strong2007, Grenier2015}. Although successful in many respects, such models often require complex parameterizations \cite{Guo2018, Gaggero2015} and extensive tuning, leading to degeneracies and limited predictive power, particularly at high energies \cite{Orlando2013,Shao:2023aoi}. Moreover, the increasing precision of observational data needs complementary approaches that can capture non-linear correlations across different wavelengths without relying on explicit assumptions. Moreover, there are high demands for neutrino astronomy, as is seen that the observations of the Galactic diffuse neutrino have a large uncertainty from incomplete theoretical models \cite{IceCube:2023ame}, due to the limitation of current high-energy observations.

Machine learning (ML) techniques have recently emerged as powerful tools for astronomical data analysis, offering the ability to capture complex nonlinear relationships in multi-wavelength datasets \cite{Feigelson2019,Smith2020,Ball2010}. These techniques have been applied to many aspects of astronomy, including the classification of astronomical objects \cite{Dubath2011, Ding:2023qxw} and regression tasks to estimate cosmological parameters \cite{Ntampaka2019}. They can uncover complex statistical associations and patterns from large-scale, high-dimensional data-patterns that are often difficult to predefine or observe directly. In this work, we construct a comprehensive, data-driven ML framework to predict gamma-ray GDE using multi-frequency radio/microwave observations from Planck PR3 \cite{Planck2018-I}. 

We employ Random Forest (RF) \cite{Breiman2001} and K-Nearest Neighbors (KNN) \cite{Cover1967} regression algorithms to establish a mapping between 9 radio frequency bands and 28 gamma-ray energy bins. Using both symmetric and asymmetric cross-validation strategies, we investigate spatial dependencies in prediction performance. Our analysis identifies the most informative spectral regions for gamma-ray prediction.

This work proves that ML can effectively capture the complex relationships between multi-wavelength GDE, providing insights that complement traditional physical modeling approaches. The methodologies developed here have broader implications for multi-wavelength astrophysical analysis and contribute to the growing field of ML applications in astronomy \cite{VanderPlas2018,Baron2019}. 

We first describe the radio and gamma-ray data, together with their processing, in Section \ref{sec:2}. Our ML methodology is presented in Section \ref{sec:3}, followed by the analysis methods in Section \ref{sec:4}. We show the results of the full-sky analysis in Section \ref{subsec:full_sky_analysis}. Finally, in Section \ref{sec:conclusion}, we give our conclusion and discussion. 

\section{Multi-wavelength observations} \label{sec:2}
Our Galaxy has been measured with multi-messenger observations, including electromagnetic waves from radio to gamma rays, cosmic rays, and neutrinos. Radio and microwave surveys have a long observational history and provide abundant data, yielding detailed maps of the Galaxy. In contrast, gamma rays at energies from hundreds of MeV up to TeV and even PeV, as well as neutrino observations, have only become accessible in recent decades \cite{Ackermann2012, LHAASO:2023gne, IceCube:2023ame}. Therefore, these observations are suitable for refining the GDE map and the current cosmic-ray theoretical framework. 

\begin{figure*}[htbp]
    \centering
    \begin{subfigure}[b]{0.48\textwidth}
        \centering
        \includegraphics[width=\linewidth]{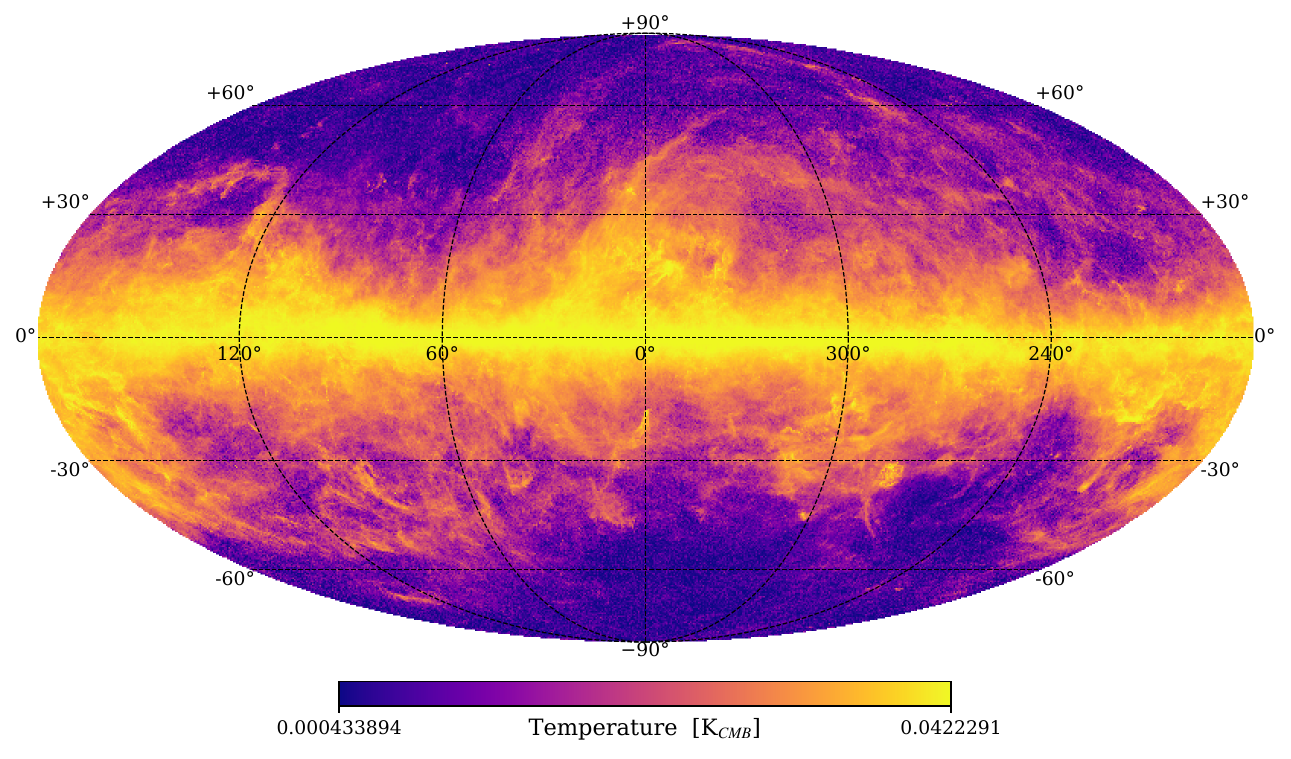} 
        \caption{Planck 353 GHz radio sky map}
        \label{fig:map_353ghz}
    \end{subfigure}
    \hfill
    \begin{subfigure}[b]{0.48\textwidth}
        \centering
        \includegraphics[width=\linewidth]{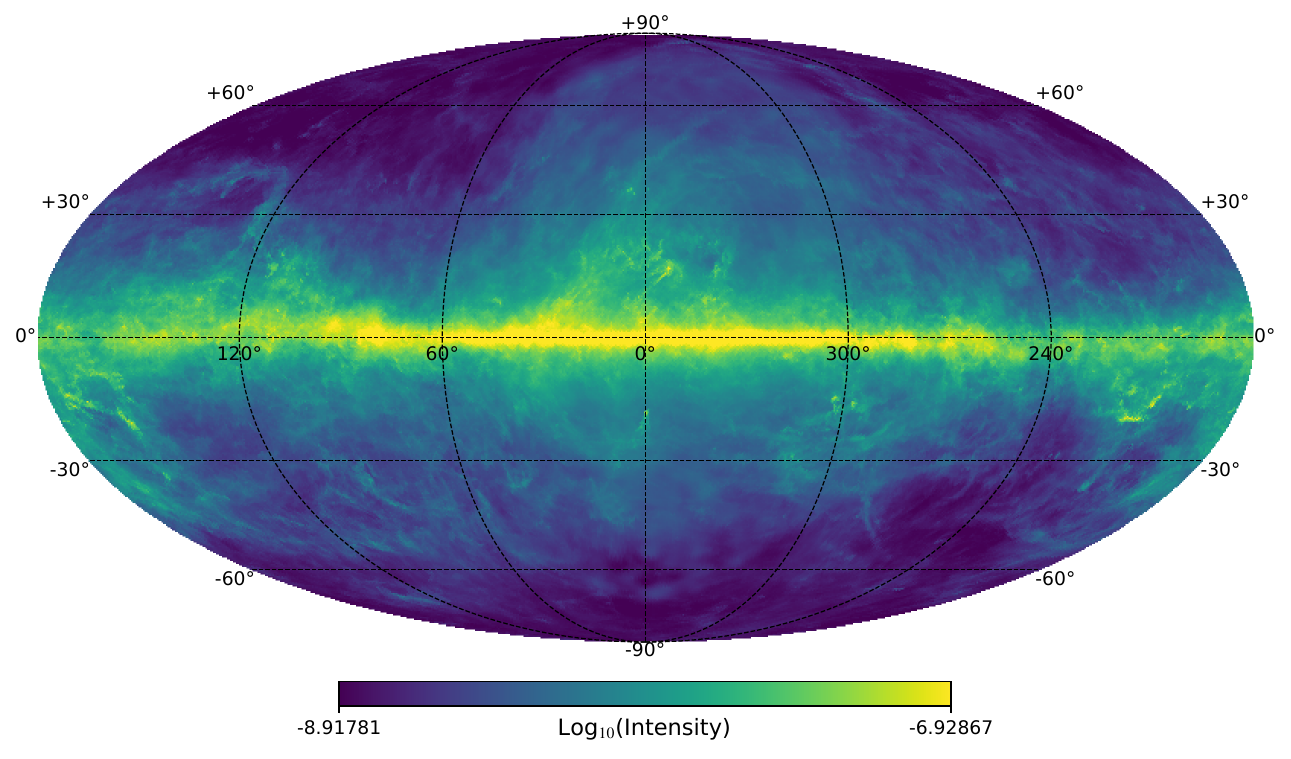} 
        \caption{Gamma-ray Galacitc diffuse emission at 687 MeV}
        \label{fig:map_fermi}
    \end{subfigure}
    
    \vspace{12pt} 
    
    \begin{subfigure}[b]{0.48\textwidth}
        \centering
        \includegraphics[width=\linewidth]{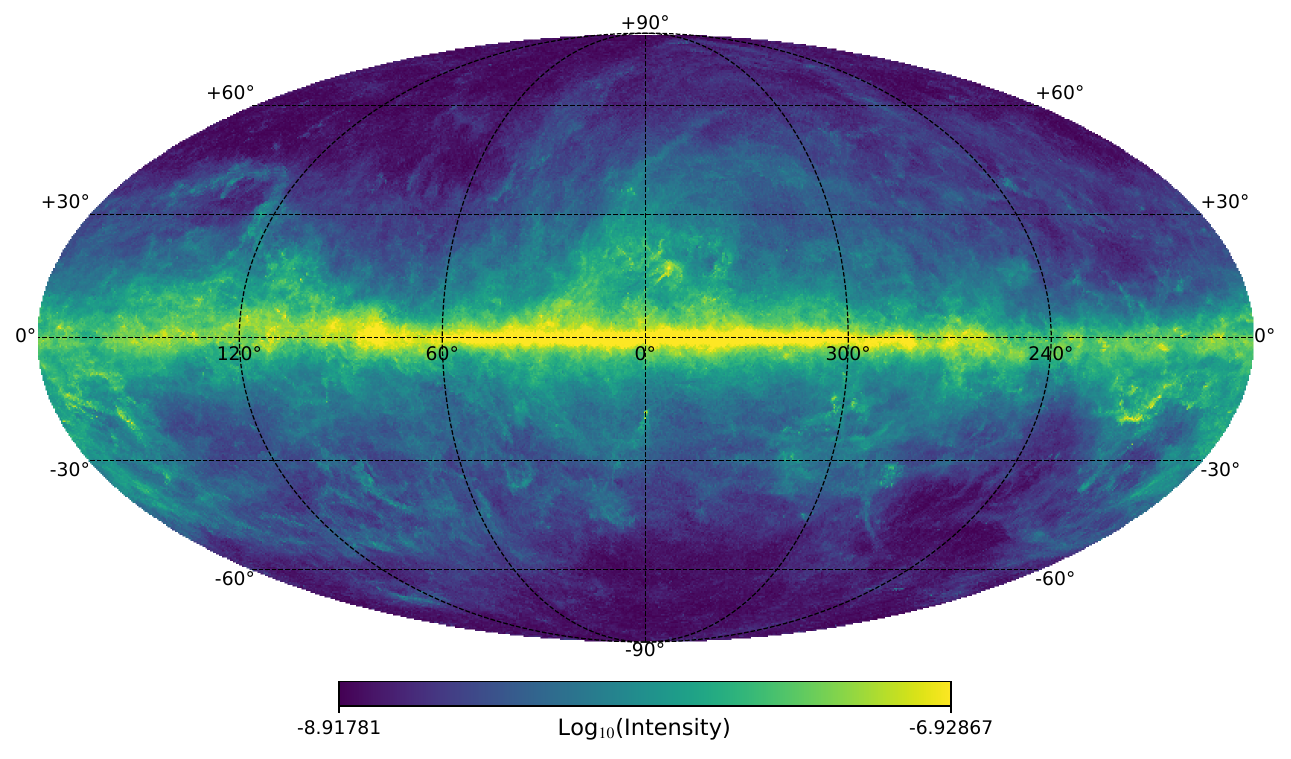} 
        \caption{ML predicted emission at 687 MeV}
        \label{fig:map_ml}
    \end{subfigure}
    \hfill
    \begin{subfigure}[b]{0.48\textwidth}
        \centering
        \includegraphics[width=\linewidth]{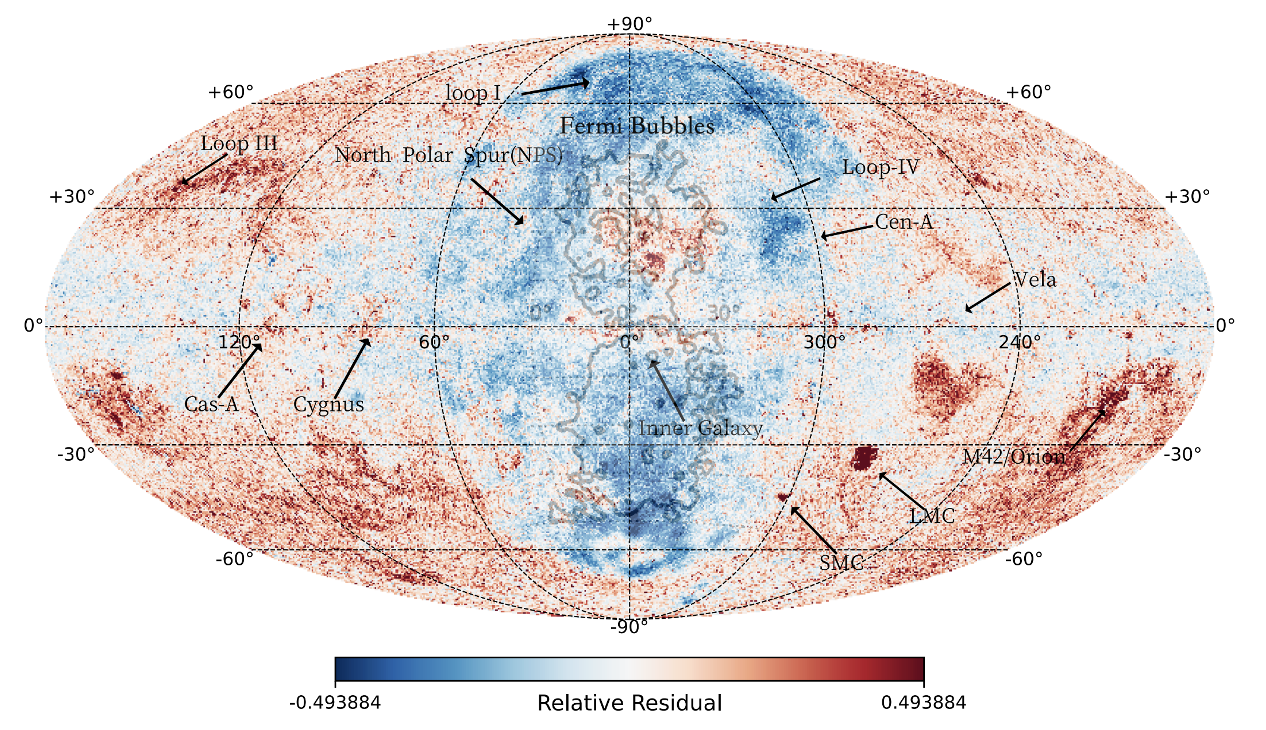} 
        \caption{Relative residual map}
        \label{fig:map_residual}
    \end{subfigure}
    
    \caption{
        (a) The radio emission map at 353 GHz \cite{Planck2015-X}. 
        (b) The \textit{Fermi}-LAT diffuse emission map at 687 MeV \cite{Fermi-LAT-IEM-2019}. 
        (c) The predicted gamma-ray map generated by the KNN model based on radio/microwave features at 687 MeV. 
        (d) The relative residual map between prediction and observation, with additional physical components or unmodeled large-scale structures labeled.
    }
    \label{fig:multiwavelength_maps}
\end{figure*}

\subsection{Planck radio and microwave data} \label{subsec:2.1}
We use full-sky radio maps from the third public data release (PR3) of the Planck Satellite, provided by the European Space Agency (ESA) \cite{Planck2018-I}, covering nine frequency bands from 30 to 857 GHz \footnote{\url{http://pla.esac.esa.int/pla/\#maps}.}. 
These maps provide a multi-frequency view of Galactic emission processes and trace different components of the ISM and the cosmic-ray environment.

To isolate physically meaningful emission components, we adopt maps processed with the COMMANDER component-separation algorithm \cite{Planck2015-VIII}. This approach combines multi-frequency observations to reconstruct astrophysical components while minimizing contamination from instrumental noise and foreground mixing \cite{Planck2018-V}. The resulting maps provide a consistent and physically interpretable representation of the diffuse sky.

Each frequency band investigates a distinct physical regime \cite{Planck2015-X,Planck2018-I}. Low-frequency channels (30–70 GHz) are dominated by synchrotron emission from electrons in Galactic magnetic fields. Intermediate frequencies (100--217 GHz) contain mixed contributions from free–free and anomalous microwave emission (spinning dust). High-frequency channels (217–857 GHz) are dominated by thermal dust emission \cite{Planck2015-X}, which serves as a robust tracer of interstellar gas.

This frequency-dependent structure is central to our analysis. By combining these bands, we construct a multi-dimensional representation of the ISM that encodes both particle and environmental information relevant to high-energy emission processes.
As an example, the thermal dust emission at 353 GHz is illustrated in Fig.~\ref{fig:map_353ghz}.

\subsection{\textit{Fermi}-LAT gamma-ray observations} \label{subsec:2.2}
The gamma-ray sky map is adopted from \textit{Fermi} diffuse emission (\texttt{gll\_iem\_v07.fits}), which provides all-sky coverage in the energy range from 50 MeV to 814 GeV \cite{Fermi-LAT-instrument,4FGL}, accessible through the Fermi Science Support Center \footnote{\url{https://fermi.gsfc.nasa.gov/ssc/data/access/lat/BackgroundModels.html}}. Instead of raw photon counts, we use the energy-binned intensity maps. This comprehensive map incorporates over a decade of observational data and has been extensively validated in the analysis of the Fourth \textit{Fermi}-LAT Source Catalog (4FGL) \cite{Acero2016, Fermi-LAT-IEM-2019, Strong2011}.

This map is a sophisticated linear combination of various physical emission components. It primarily comprises: (1) gas-related emission maps derived from H~{\sc i} (atomic hydrogen) and CO (tracing molecular hydrogen) surveys; (2) an Inverse Compton (IC) component modeled via GALPROP, representing interactions between cosmic-ray electrons and the interstellar radiation field; (3) a template for the Dark Neutral Medium (DNM), or dark gas, which captures interstellar gas not traced by H~{\sc i} or CO lines; and (4) a model for unresolved Galactic sources. Crucially, the DNM component is based on dust optical depth maps from the Planck satellite \cite{Planck2015-X,Acero2016,Fermi-LAT-IEM-2019}. This intrinsic dependence of the gamma-ray GDE on Planck dust data provides strong physical justification for our ML approach, reinforcing the validity of using multi-frequency microwave/radio observations to predict gamma-ray emission. Furthermore, to account for large-scale excesses not fully explained by gas or IC templates, the model incorporates empirical patch components derived from residuals, which correspond to specific structures such as Loop~I and Fermi Bubbles \cite{Fermi-LAT:FB}.

To ensure compatibility with the radio/microwave maps, all gamma-ray maps are projected onto a common HEALPix grid with matching angular resolution. This alignment allows for a direct pixel-by-pixel comparison and enables the construction of a supervised mapping between low-energy tracers and high-energy emission. A representative reference map of gamma-ray GDE at 687 MeV is shown in Fig.~\ref{fig:map_fermi}.

\subsection{Data processing} \label{subsec:2.3}
We integrate multi-wavelength observations to construct a unified dataset for ML analysis. Our data processing pipeline homogenizes the spatial grids of different surveys and extracts relevant features to establish a direct mapping from radio and microwave tracers to gamma-ray emission.

We standardize all maps with the \textsc{HEALPix} pixelization scheme to minimize distortion at high latitudes, setting the pixel size to approximately $\mathbf{0.125^\circ}$ to match the native granularity of the gamma-ray observations. For each \textsc{HEALPix} pixel center $(\theta, \phi)$, we compute the corresponding coordinates in the CAR system and perform pixel-wise mapping to extract the gamma-ray intensity values.
    
We aim to predict gamma-ray emission solely on the basis of radio and microwave tracers. For each of the 28 \textit{Fermi}-LAT energy bins, we construct an independent regression task. The input feature space $\mathbf{X}$ for every sampled pixel consists exclusively of physical values from the nine Planck frequency bands (30--857 GHz).

The unified data aligns multi-wavelength observations on a unified spherical grid ($\mathbf{X} \in \mathbb{R}^9 \to \mathbf{I_{\gamma}} \in \mathbb{R}^1$), forming the basis for a data-driven investigation of the relationship between ISM tracers and gamma-ray emission.   

\section{Machine learning methods} \label{sec:3}
To establish a mapping between radio and gamma-ray emission, we seek a function $I_\gamma = f(I_{\rm radio})$, where $I_{\rm radio}$ represents the multi-frequency Planck observations and $I_\gamma$ denotes the gamma-ray intensity in a given energy bin.

Rather than constructing this mapping from first-principles physical modeling, we adopt a data-driven approach. This allows us to capture non-linear correlations between observables without imposing strong prior assumptions about cosmic-ray propagation or emission mechanisms.

Importantly, the function $f$ should not be interpreted as a purely mathematical object. Instead, it represents an empirical encoding of the dominant physical relationship between ISM, radiation fields, and high-energy processes in the Galaxy.

\subsection{Machine learning framework} \label{subsec:ml_models}
We employ supervised ML to approximate the mapping $f$. The input features consist of the nine Planck frequency maps, whereas the target variable is the gamma-ray intensity in a specific energy bin.

We apply two well-established algorithms, RF regression \cite{Breiman2001} and KNN regression \cite{Cover1967}. These algorithms are implemented using the scikit-learn library \cite{Pedregosa2011}, adhering to best practices in astro-statistics \cite{Feigelson2019}.

Random Forest regression constructs an ensemble of decision trees, reducing overfitting through bootstrap aggregation and feature randomization. It is particularly valuable \cite{Ding:2023qxw} for its ability to rank the importance of characteristics and identify which radio bands physically drive gamma-ray emission. The KNN regression predicts target values based on the local structure of the feature space and is effective in capturing morphological similarities in diffuse emission.

Both algorithms demonstrate excellent performance, with high coefficients of determination ($R^2 > 0.95$ in optimal bands), where $R^2$ is defined as
\begin{equation}
    R^2 = 1 - \frac{\sum_{i=1}^{N} \left(\log_{10}I_{i, \rm data} - \log_{10}I_{i, \rm pred} \right)^2}{\sum_{i=1}^{N} \left( \log_{10}I_{i, \rm data} - \overline{\log_{10}I_{\rm data}} \right)^2}.
    \label{equ:R2}
\end{equation}

Here $I_{i, \rm data}$, and $I_{i, \rm pred}$ are gamma-ray data and prediction, respectively, for the $i$-th pixel. And $\overline{\log_{10}I_{\rm data}}$ are the averaged data.
We also perform a comprehensive model comparison including Linear Regression (LR) and Artificial Neural Networks (ANNs). However, these alternatives do not yield significant performance gains compared to RF and KNN, often struggling to balance computational efficiency with the ability to capture complex non-linearities \cite{Smith2020}. Consequently, we focus our analysis on RF and KNN. Moreover, as tested with all the cases in this study, RF and KNN give very close results. Therefore, we primarily present results from the KNN model as it has a higher computational efficiency.

With the preprocessed data described in Sect.~\ref{subsec:2.3}, we adopt an independent regression strategy. Each of the 28 gamma-ray maps employs a separate model trained on the same nine radio frequency bands. This enables the algorithms to learn energy-dependent physical correlations without imposing rigid spectral constraints across the gamma-ray band.

\subsection{Training strategy}
To evaluate the physical robustness of the learned mapping, we adopt multiple training strategies that go beyond standard random splitting.

First, we perform cross-hemisphere validation, where the model is trained on one Galactic hemisphere and tested on the other, probing the large-scale transferability of the mapping and tests whether it reflects universal Galactic properties.

Second, we implement high-resolution spatial segmentation, dividing the sky into smaller regions, and evaluating the performance of cross-region prediction. This allows us to determine whether the mapping is governed by spatial proximity or by intrinsic emission conditions.
Moreover, it ensures that model performance is not driven by trivial interpolation, but instead reflects meaningful physical relationships\cite{Feigelson2019, Ivezic2014}.

Finally, we perform the full-sky analysis with spatially binned features to assess the large-scale prediction of the radio-gamma mapping. 

The performance of the model is evaluated by the coefficient of determination ($R^2$), mean squared error (MSE) and mean absolute relative error (MARE), capturing the explanation of variance and both precision of absolute and relative prediction\cite{Chai2014, Willmott2005}.

However, the evaluation metrics are not interpreted purely as statistical indicators. Instead, variations in performance across energy, frequency, and spatial configurations are used as probes of the underlying physical processes governing diffuse emission.

\subsection{Residual analysis and uncertainty estimates}
\label{subsec:residual_and_uncertainty}

To identify deviations from the learned mapping, we calculate the relative residual defined as
\begin{equation}
    \delta_{\mathrm{rel}} = \frac{I_{\mathrm{pred}} - I_{\mathrm{data}}}{I_{\mathrm{data}}},
    \label{equ:residual}
\end{equation}
where $I_{\mathrm{pred}}$ and $I_{\mathrm{data}}$ are the predicted and observational gamma-ray data. These residuals highlight regions where the dominant radio-gamma correlation breaks down. Such deviations can arise from additional emission components, environmental variations, or limitations in the construction of the reference diffuse model \cite{Acero2016, Fermi-LAT-IEM-2019}. In this sense, the residual map provides a data-driven diagnostic for identifying non-standard structures in the gamma-ray sky.

To quantify the robustness of the ML methods, we obtain the corresponding uncertainty. Within the KNN framework, the relative uncertainty is dynamically derived from the local variance of its nearest K neighbors in the feature space \cite{Hastie2009, Cover1967}. It is defined as the ratio of the standard deviation ($\sigma$) to the mean intensity ($I_{\mathrm{mean}}$) of these neighbors,
\begin{equation}
    U_{\mathrm{rel}} = \frac{\sigma}{I_{\mathrm{mean}}}.
\end{equation}

\begin{figure*}[htbp]
    \centering
    \begin{subfigure}[t]{0.48\textwidth}
        \centering
        \includegraphics[width=\linewidth]{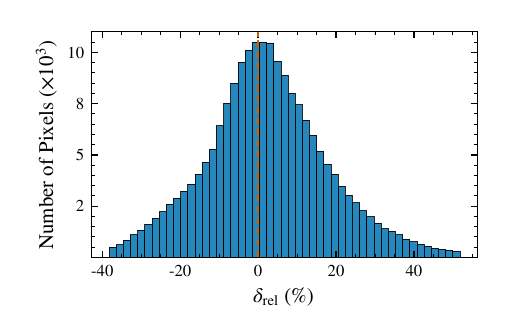}
        \caption{Distribution of relative residuals $\delta_{\mathrm{rel}}$ at 687~MeV.}
        \label{fig:hist_residual}
    \end{subfigure}
    \hfill
    \begin{subfigure}[t]{0.48\textwidth}
        \centering
        \includegraphics[width=\linewidth]{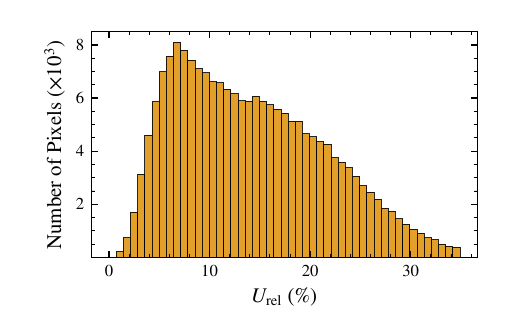}
        \caption{Distribution of relative uncertainties $U_{\mathrm{rel}}$ at 687~MeV.}
        \label{fig:hist_uncertainty}
    \end{subfigure}

    \vspace{12pt}

    \begin{subfigure}[t]{0.48\textwidth}
        \centering
        \includegraphics[width=\linewidth]{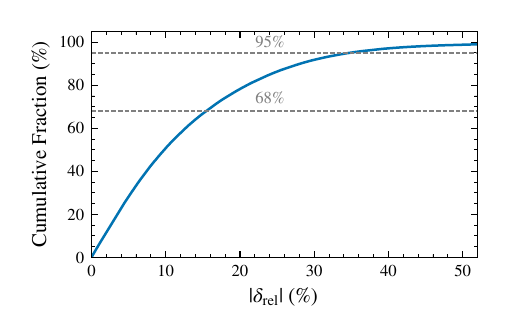}
        \caption{Cumulative distribution of $|\delta_{\mathrm{rel}}|$.}
        \label{fig:cdf_residual}
    \end{subfigure}
    \hfill
    \begin{subfigure}[t]{0.48\textwidth}
        \centering
        \includegraphics[width=\linewidth]{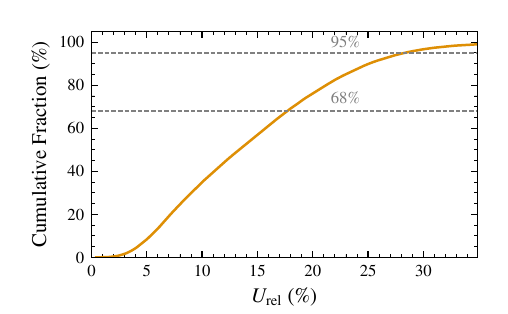}
        \caption{Cumulative distribution of $U_{\mathrm{rel}}$.}
        \label{fig:cdf_uncertainty}
    \end{subfigure}

    \caption{
        Statistical properties of residuals and uncertainties for the KNN prediction of full-sky gamma-ray GDE at 687~MeV.
        (a)~Distribution of relative residuals.
        (b)~Distribution of relative uncertainties.
        (c)~Cumulative distribution of absolute residuals.
        (d)~Cumulative distribution of relative uncertainties.
    }
    \label{fig:residual_uncertainty_summary}
\end{figure*}

As an example, we present one of the full-sky results here to illustrate the reliability of the ML method (general discussions will be presented in Section~\ref{subsec:full_sky_analysis}). We chose the 687~MeV energy band as a representative case, as the ML model achieves its optimal performance at this energy, producing an overall $R^2 = 0.96$. Figure~\ref{fig:residual_uncertainty_summary} shows the residual and uncertainty statistics at 687~MeV. Figure \ref{fig:hist_residual} shows the distribution of relative residuals $\delta_{\mathrm{rel}}$. It has a mean of $+2.79\%$ and a median of $+1.76\%$, indicating a slight global overestimation by the model. The standard deviation of $18.57\%$ and the 95\% containment interval $[-29.65\%, +40.18\%]$ confirm that the majority are well predicted, while the extended tails capture localized astrophysical anomalies. Figure \ref{fig:hist_uncertainty} presents the distribution of relative uncertainties. It has a peak around $6.5\%$, with a mean of $14.42\%$ and a median of $13.33\%$. This distribution shows that most predicted regions are reliable with the KNN model, whereas high-uncertainty pixels are due to low-flux or structurally complex.

Figures \ref{fig:cdf_residual} and \ref{fig:cdf_uncertainty} show the cumulative distributions of $|\delta_{\mathrm{rel}}|$ and $U_{\mathrm{rel}}$, respectively. The Cumulative Distribution Function (CDF) of the absolute residuals reaches 68\% and 95\% at $15.45\%$ and $34.71\%$, respectively, quantifying the global precision of the mapping. For relative uncertainty, the CDF reaches 68\% and 95\% at $17.71\%$ and $28.28\%$, respectively. Within 68\%, the residual CDF is smaller than that of uncertainty, indicating that the variance of predicted values is well within the uncertainty. 

Moreover, we perform k-fold cross-validation by partitioning the dataset into independent subsets and evaluating the dispersion of predictions in multiple folds \cite{Hastie2009}. The resulting maps show a similar distribution. It tells the spatial variance does not bring additional uncertainty in this study.

These analyses present that our framework is remarkably robust against variations in training data and model realization, confirming that the identified residual structures are statistically significant and physically meaningful.

\section{Analysis} \label{sec:4}

\subsection{Cross-Hemisphere Validation} \label{subsec:symmetric_validation_physical}
To investigate whether the radio–gamma correlation is a local relation or a globally transferable property of the Galaxy, we performed a cross-hemisphere validation. Rather than adopting a random train–test split, it investigates whether a learned mapping in one large-scale Galactic environment can generalize to its symmetric counterpart \cite{Feigelson2019, Blaauw1960}. In this sense, the test provides not only a measure of predictive accuracy but also a physically motivated probe of large-scale structural symmetry in GDE.

To clarify directional asymmetry with respect to the Galactic Center (GC), we define the Eastern Hemisphere (EH) as $l \in (0^\circ, +180^\circ]$ and the Western Hemisphere (WH) as $l \in (180^\circ, 360^\circ]$. We train in one hemisphere and test in the other, allowing us to evaluate whether the empirical radio–gamma mapping is universal or environment-dependent.

\begin{figure}[htbp]
    \includegraphics[width=0.45\textwidth]{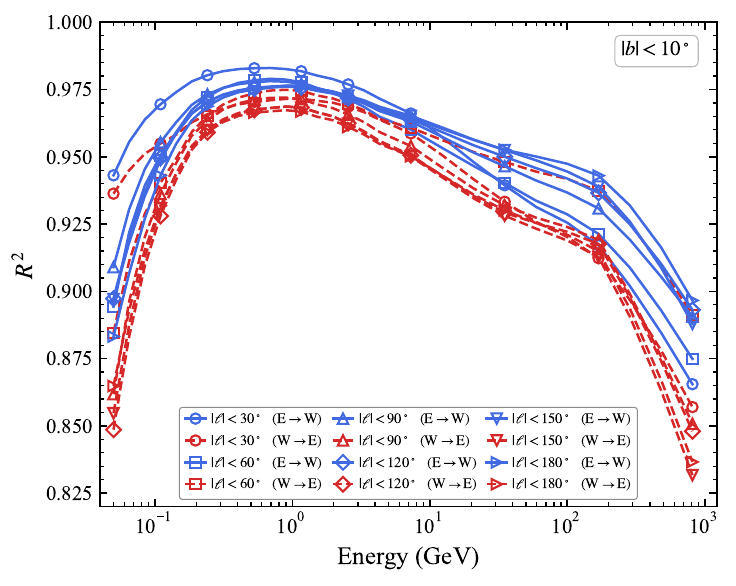}
    \caption{
    The $R^2$ for symmetric cross-hemisphere validation.
    Models trained on EH and tested on WH are represented in blue solid lines ($E \to W$). Here, the parameter $\theta$ is set as 30, 60, 90, 120, 150 and 180.
    The inverse configuration is represented by red dashed lines ($W \to E$). }
    \label{fig:ml_performance_comparison}
\end{figure}

We first examined how the predictive relation depends on the spatial extent of the training data using a cumulative region-growing strategy. Symmetric regions of interest (ROIs) are defined within the Galactic plane ($|b| < 10^\circ$), with the longitudinal extent $\theta$ increasing from $30^\circ$ to $180^\circ$ in steps of $30^\circ$. For each $\theta$, we perform a bidirectional validation, East-to-West ($E \to W$), where the model is trained on EH and tested on WH, and West-to-East ($W \to E$), the inverse configuration.

This setup is physically informative because the dependence indicates whether the learned mapping is controlled mainly by local structures or by large-scale statistical regularities \cite{Carilli2004}. If model performance remains stable as $\theta$ varies, the radio–gamma relation is likely governed by global Galactic properties rather than by narrowly localized features.

As shown in Fig.~\ref{fig:ml_performance_comparison}, the performance is energy dependent, with high values of $R^2$ occurring in the 0.1 to 10 GeV range. In this energy range, gamma-ray emission in the Galactic plane is expected to be dominated by hadronic interactions, in which cosmic-ray protons collide with the interstellar gas and produce neutral pions \cite{Strong2007, Ackermann2012}. Where the gamma-ray intensity exhibits a linear correlation with the gas column density \cite{Fermi-LAT:2009pgs}. Since Planck high-frequency channels effectively trace dust-associated gas, the strong agreement between prediction and observation indicates that the dominant target material for gamma-ray production is well encoded in the radio/microwave data \cite{Planck2015-X,Su2010}.

Equally important is the weak dependence of performance on the size of the training region $\theta$, as shown in Fig.~\ref{fig:ml_performance_comparison}. The stability of the results (for each energy, $R^2$ changes not more than 6\%) over different spatial extents implies that the learned mapping is not restricted to small local structures but instead reflects a robust large-scale coupling between ISM tracers and gamma-ray emission across the Galactic disk.

A small but systematic asymmetry can be seen between the two transfer directions, with $E \to W$ predictions consistently performing slightly better than $W \to E$. Although the difference is modest, its persistence across energy bins suggests that the two hemispheres are not statistically equivalent. This effect is consistent with the presence of additional structural complexity in the inner EH of the Galaxy, which can be seen in Fig.~\ref{fig:galactic_flux_distribution}. It is potentially associated with spiral-arm tangents or bar-related features, which may provide a richer training environment and, therefore, improve transferability.

To determine whether the hemispheric asymmetry identified above is a global property or instead driven by specific regions, we repeated the analysis using discrete, non-overlapping spatial windows that exclude the GC. It isolates the contribution of different Galactic environments and removes the cumulative influence of the innermost regions, where line-of-sight confusion and structural complexity are strongest.

\begin{figure}[htbp]
 
    \includegraphics[width=0.45\textwidth]{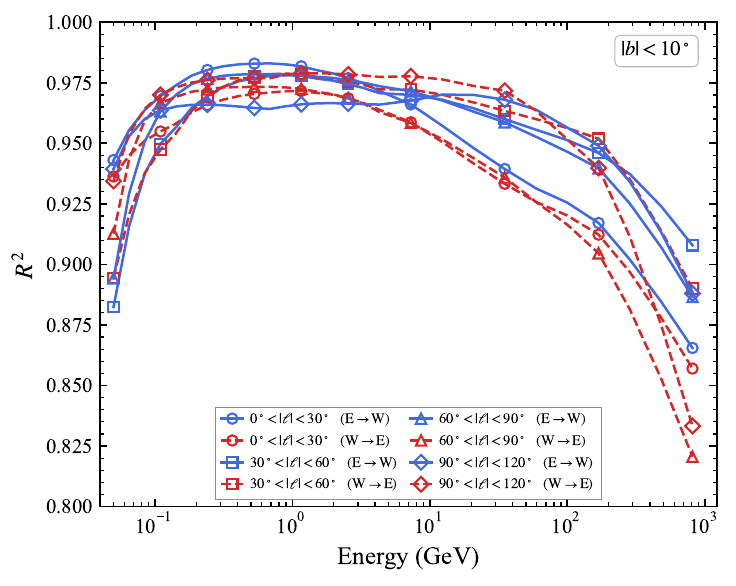}
    \caption{
        The $R^2$ for models trained on discrete spatial windows of $30^\circ$ width, excluding the GC.
        Training on the EH and testing on the symmetric WH is represented by blue solid lines ($E \to W$).
        The inverse configuration is represented by red dashed lines ($W \to E$).
        The parameter $\theta$ denotes the angular offset from the GC.
    }
    \label{fig:non_symmetric_ml_performance}
\end{figure}

The results of this segmented analysis shown in Fig.~\ref{fig:non_symmetric_ml_performance} show that the systematic performance gap is significantly reduced in spatially segregated regions. The $R^2$ for both directions become largely comparable across most energy bands, except the region $60^\circ< |l| < 90^\circ$, which represents a spiral arm transition zone composed of multiple small-scale features rather than a single massive structure \cite{Reid:2014boa, Bland-Hawthorn:2016lwg}.
This suggests that the observed directional asymmetry is primarily localized to the inner Galaxy ($|l| < 30^\circ$), likely driven by structural complexities such as the Galactic bar and the spiral arm tangents \cite{Benjamin2005,Reid2019}, while the outer disk exhibits a more homogeneous distribution that is robust to the training direction.

Having established the robustness of the large-scale mapping, we next examine which radio frequency ranges carry the most physically relevant information for predicting gamma-ray emission. It directly probes which emission components in the radio/microwave sky are most closely linked to the processes responsible for diffuse gamma rays.

\begin{figure}[htbp]
 
    \includegraphics[width=0.45\textwidth]{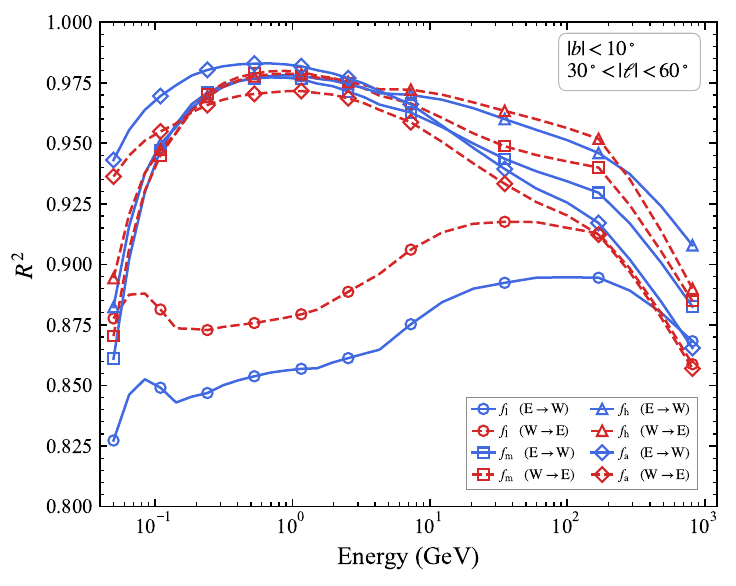}
    \caption{
        Performance comparison of models using different frequency band groups as features. 
        The analysis works on Inner Disk region: East ($l \in [30^\circ, 60^\circ]$) and West ($l \in [-60^\circ, -30^\circ]$).
        Solid blue lines ($E \to W$) represent training on the EH and testing on the WH.
        Dashed red lines ($W \to E$) represent training on the WH and testing on the EH.
        Input features are grouped into, low-frequency (30--70 GHz), intermediate (100--217 GHz), high-frequency (353--857 GHz), and all bands combined.
    }
    \label{fig:frequency_band_performance}
\end{figure}

As shown in Fig.~\ref{fig:frequency_band_performance}, the high-frequency Planck bands, dominated by thermal dust emission, achieve predictive performance that is almost comparable to that obtained using all nine frequency bands, especially in the energy band of 0.1 -- 10 GeV, where low-frequency channels perform significantly worse. This indicates that the most informative predictors of gamma-ray emission of this range are those that trace interstellar gas and dust. This is because $I_{\gamma} \propto N_{gas} \times q_{CR}$, where $I_\gamma$ is the intensity of gamma-ray emission, $N_{gas}$ is the column density of gas, and $q_{CR}$ is the cosmic-ray emissivity \cite{Strong2007, Draine2011}.

The result provides strong empirical support for a hadronically dominated origin of Galactic diffuse gamma rays in the 0.1 to 10 GeV energy range. In such a scenario, the thermal dust emission serves as an effective proxy for the target material \cite{Planck2015-X}, and the Galactic accelerators, such as Supernova Remnant, inject high-energy cosmic rays. For energies greater than 10 GeV, the predictive power of low-frequency bands tends to increase. This supports a leptonic origin of gamma rays, revealing the native connection between synchrotron radiation and IC. 

Instrumental factors may further enhance this contrast. In particular, the lower angular resolution of the low-frequency maps reduces their ability to capture fine morphological structures relative to that of the high-frequency channels. However, this observational limitation does not alter the main conclusion that dust-associated tracers remain substantially more informative than synchrotron-dominated tracers for reconstructing diffuse gamma-ray emission.

\subsection{High-resolution spatial segmentation} \label{subsec:high_res_flux_analysis}
We now move to a more localized analysis of the Galactic plane. The goal is to determine which property of a region controls the model's ability to generalize, spatial proximity, or similarity in the intrinsic emission environment. This distinction is important because intuition might suggest that nearby regions should be easier to predict, whereas a physically driven mapping may instead depend on shared local conditions such as gas density, radiation fields, and cosmic-ray content.

\begin{figure}[htbp]
    \includegraphics[width=0.5\textwidth]{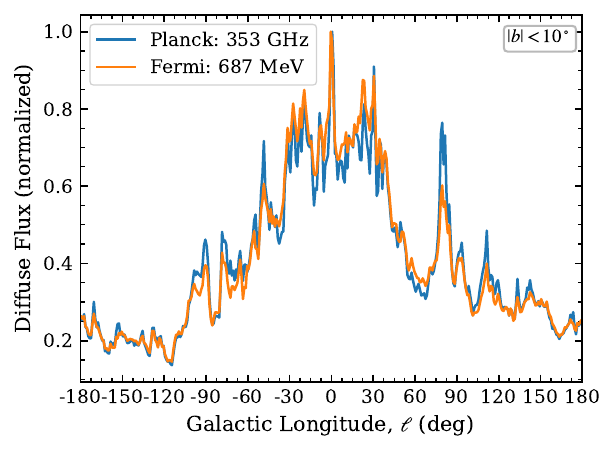}
    \caption{Normalized longitudinal profiles of diffuse Galactic emission along the Galactic plane ($|b| < 10^\circ$). The blue solid line shows radio emission at 353~GHz, while the orange solid line shows diffuse gamma-ray intensity at 687~MeV. Both profiles are normalized to their respective maxima at the GC ($\ell = 0^\circ$).
    }
    \label{fig:galactic_flux_distribution}
\end{figure}

Figure~\ref{fig:galactic_flux_distribution} shows the normalized longitudinal profiles of 353 GHz radio and 687 MeV gamma-ray intensity along the Galactic plane. The broad similarity of the two profiles confirms a strong global correspondence. At the same time, noticeable local deviations are presented, indicating the absence of nonthermal particles or leptonic origin. These variations motivate a more fine-grained regional study.

To quantify this effect while minimizing the confounding influence of the sparse outer disk, we restricted the analysis to the inner Galaxy ($|l| < 30^\circ$) and a narrow latitudinal strip ($|b| < 1^\circ$). This region is divided into 60 contiguous longitudinal subregions of width $1^\circ$ each: 58 outer bins with $1^\circ \leq |l| < 30^\circ$ and 2 centermost bins with $|l| < 1^\circ$. We then perform a cross-validation from all over between these 60 subregions using KNN regression, with all nine Planck maps as input and the 687~MeV gamma-ray band as output. 

For any given pair of subregions $i$ and $j$, we train a model exclusively on the pixels of region $i$ to predict the gamma-ray intensity in region $j$. To prevent regions with extreme brightness from dominating the loss function, the directional Mean Squared Error ($\text{MSE}_{i \to j}$) is computed,
\begin{equation}
    \text{MSE}_{i \to j} = \frac{1}{N_j} \sum_{k=1}^{N_j} \left( \log_{10} I_{\text{pred}, k}^{(j)} - \log_{10} I_{\text{data}, k}^{(j)} \right)^2,
\end{equation}
where $N_j$ is the total number of valid pixels in the testing region $j$, and the index $k$ is the $k-$th pixel within this region. Here, $I_{\text{pred}, k}^{(j)}$ and $I_{\text{data}, k}^{(j)}$ denote the predicted and observed gamma-ray intensities at the $k$-th pixel, respectively. To ensure a symmetric and unbiased evaluation of the environmental mismatch between the two regions, we compute the bi-directional average prediction error as follows,
\begin{equation}
    \overline{\text{MSE}}_{ij} = \frac{1}{2} \left( \text{MSE}_{i \to j} + \text{MSE}_{j \to i} \right).
\end{equation}

\begin{figure}[htbp]
    \centering
    \includegraphics[width=0.9\linewidth]{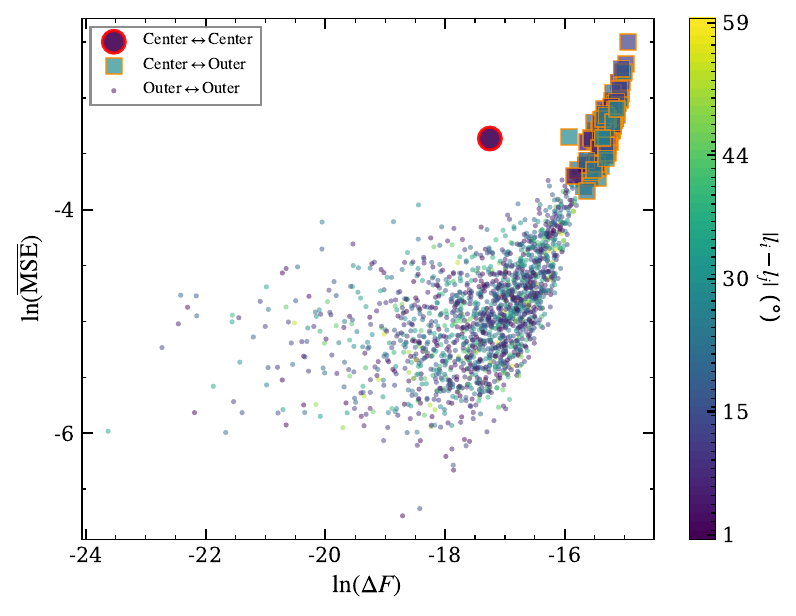}
    \caption{
        Cross-region prediction performance at 687~MeV for all 60 longitudinal subregions within $|b| < 1^\circ$ and $|l| < 30^\circ$. 
        Dots ($\mathrm{Outer} \leftrightarrow \mathrm{Outer}$) are pairs among the 52 outer bins ($1^\circ \leq |l| < 30^\circ$), squares ($\mathrm{Center} \leftrightarrow \mathrm{Outer}$) present pairs between the GC bin ($|l| < 1^\circ$) and each outer bin, and circle ($\mathrm{Center} \leftrightarrow \mathrm{Center}$) show the single pair formed by the two adjacent center bins themselves.
        The color bar encodes the longitudinal separation $|l_i - l_j|$.
    }
    \label{fig:center_vs_outer_analysis}
\end{figure}

We explore the relationship between the absolute flux difference $\Delta F_{ij} = |F_i - F_j|$ (where $F$ is the mean flux of the region) and the corresponding prediction error $\ln(\overline{\text{MSE}}_{ij})$, as shown in Fig.~\ref{fig:center_vs_outer_analysis}. This tests whether predictive failure is driven by geometric spatial separation or by discrepancies in intrinsic physical emission.

In Fig.~\ref{fig:center_vs_outer_analysis}, the dots represent pairs drawn from the 58 outer bins ($|l| \geq 1^\circ$). The squares denote pairs in which one is a GC bin ($|l| < 1^\circ$) and the other lies among the outer bins. The circle corresponds to the pair formed by the two adjacent center bins. It is clear that region pairs with similar fluxes generally yield low errors, even when they are widely separated. In contrast, spatially adjacent regions can produce poor predictions if their intrinsic fluxes differ substantially. This behavior demonstrates that the model generalizes primarily across regions with similar physical conditions, rather than those that are geographically adjacent.

It is interesting that the center-center pair has an anomalously high prediction error, even though they are adjacent and have nearly identical mean intensities. The mean fluxes are almost identical for the center and nearby outer pairs, a KNN mapping trained on one center bin fails to reproduce others. It indicates that structural complexity, not just flux mismatch, drives predictive failure in the GC \cite{Fermi-diffuse-model-2018}.

We further extended this to a two-dimensional segmentation of the inner Galactic plane ($|l| < 30^\circ, |b| < 5^\circ$), partitioning the region into 600 subregions to include latitudinal variations. In this test, the same trend persists. Regions with intermediate flux levels, close to the global median, tend to yield the lowest cross-region prediction errors. The GC is even more special, where its structural complexity leads to poorer learning performance. This broader result supports the interpretation that good transferability is associated with astrophysical homogeneity rather than with simple spatial closeness.

\section{Results and implications} \label{subsec:full_sky_analysis}

Having established that the radio–gamma mapping is both statistically robust and physically interpretable, we apply the framework to the full sky. 

\subsection{Full-sky gamma-ray prediction}
The data set is randomly divided into 80\% training samples and 20\% testing samples. 
The full-sky predictions for 687 MeV are shown in Fig.~\ref{fig:map_residual} (other energy bins generally behave similarly). The model successfully reproduces the major large-scale morphological features of the diffuse gamma-ray sky over a broad energy range. 

The residual maps highlight regions where the radio-gamma correlation is insufficient, thus indicating the presence of additional emission processes, environmental features, or limitations in the construction of the reference diffuse model. As shown in Fig.~\ref{fig:map_residual}, positive residuals (prediction over-data) are prominently seen toward Vela, Cen A, Loop III, and the Large and Small Magellanic Clouds (LMC/SMC). This mismatch is physically intuitive, as the Planck maps trace the abundant gas and dust content in these systems, including nearby galaxies, prompting the learned ML relation to predict an associated hadronic gamma-ray signal. However, in \textit{Fermi} GDE maps, these specific sources are often treated separately from the Galactic background emission \cite{Acero2016, Fermi-LAT-IEM-2019}. This naturally yields positive residuals, demonstrating that the ML model is highly sensitive to matter structures present in the radio data regardless of predefined template boundaries.
Conversely, negative residuals (prediction under-data) are most striking in the Loop I and North Polar Spur (NPS) regions. This suggests the presence of an additional inverse-Compton-dominated or locally enhanced high-energy electron component not adequately encoded in Planck inputs \cite{Dickinson2018}. Furthermore, the adopted \textit{Fermi} GDE model itself includes empirical patch components in this region to force a fit \cite{Ackermann2012}. Therefore, part of the discrepancy may also reflect systematic uncertainties in the reference gamma-ray model. The residual is meaningful regardless of which of these contributions dominates.

More generally, regions with negative residuals may host excess high-energy processes, such as unresolved inverse-Compton structures or unresolved source populations \cite{4FGL}. Positive residuals may indicate regions where the GDE model has reassigned emission components relative to the matter distribution traced by radio data. The residual map serves as a physically informative diagnostic for identifying non-standard structures and for guiding future multi-messenger investigations. Such as the positive residual in the north Fermi bubble may imply a hybrid origin of this structure. 

\begin{figure}[htbp]
    \centering
    \includegraphics[width=0.9\linewidth]{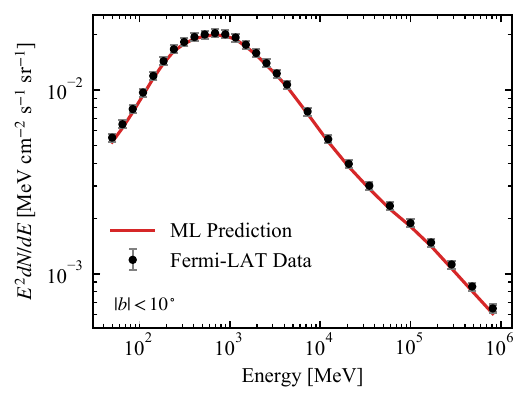}
    \caption{
        ML predicted (red solid line) and observed (black dots) SED of gamma-ray GDE within the inner Galactic plane ($|b| < 10^\circ$) are shown. 
    }
    \label{fig:sed_reconstruction_inner}
\end{figure}

Figure~\ref{fig:sed_reconstruction_inner} shows the spectral energy distribution (SED) of GDE in the inner Galactic plane with $|b| < 10^\circ$. 
The SED agreement implies that the multi-frequency radio inputs encode information about the physical environments that shape the gamma-ray spectrum. In particular, the result suggests that the spatially varying coupling between gas, dust, radiation fields, and cosmic rays is sufficiently regular that a data-driven model can recover the broad spectral form.

We also find that the inclusion of neighboring gamma-ray energy bands as auxiliary inputs can reduce residuals in complex regions such as Loop I. It is reasonable as the adjacent high-energy channels share the continuum emission information. By combining radio, gamma-ray, and potentially X-ray tracers, one may construct a more complete empirical representation of Galactic non-thermal emission across multiple messengers, such as cosmic ray and neutrino \cite{IceCube:2023ame, LHAASO:2023gne, Shao:2023aoi}.

\subsection{Physical Modeling vs. Machine Learning}
\label{sec:benchmark}
To comprehensively evaluate the predictive capability of our ML approach, we compare it with the results of GALPROP \cite{Shao:2023aoi}. 
To quantify the absolute flux deviation, we calculate the Mean Absolute Relative Error (MARE), defined as,
\begin{equation}
    MARE = \frac{1}{N} \sum_{i=1}^{N} \left| \frac{I_{i, \rm pred} - I_{i, \rm data}}{I_{i, \rm data}} \right| \times 100\%,
\end{equation}
where $I_{i, \rm data}$ and $I_{i, \rm pred}$ represent the observed intensity and the predicted intensity on the $i$-th pixel, respectively.

\begin{table*}[htbp]
    \centering
    \caption{The $R^2$ and MARE of GALPROP physical model \cite{Shao:2023aoi} and the KNN ML model across six distinct Galactic regions at $\sim$4.3 GeV. In this analysis, the disk ($|b| \leq 10^\circ$) is partitioned by Galactic longitude ($|l|$), and halo regions are defined at $|b| > 20^\circ$.}
    \label{tab:evaluation_metrics}
    \begin{tabular}{lcccc}
        \toprule
        \multirow{2}{*}{Galactic Region} & \multicolumn{2}{c}{GALPROP (Physical Model)} & \multicolumn{2}{c}{KNN Model (Data-Driven)} \\
        \cmidrule(lr){2-3} \cmidrule(lr){4-5}
        & $R^2$ & MARE (\%) & $R^2$ & MARE (\%) \\
        \midrule
        All Sky                                                                 & 0.8931 & 21.17 & 0.9347 & 21.18 \\
        Zone 1: North Halo ($b > 20^\circ$)                                     & 0.5908 & 25.14 & 0.8192 & 21.10 \\
        Zone 2: South Halo ($b < -20^\circ$)                                    & 0.6372 & 20.19 & 0.6923 & 27.30 \\
        Zone 3: Inner Disk \& GC ($|b| \leq 10^\circ, |l| \leq 60^\circ$)        & 0.8624 & 22.26 & 0.9465 & 14.70 \\
        Zone 4: Outer Disk I ($|b| \leq 10^\circ, 60^\circ < l \leq 180^\circ$) & 0.9474 & 10.91 & 0.9407 & 11.30 \\
        Zone 5: Outer Disk II ($|b| \leq 10^\circ, -180^\circ \leq l < -60^\circ$) & 0.9301 & 15.18 & 0.9592 & 10.21 \\
        \bottomrule
    \end{tabular}
\end{table*}

As shown in Table \ref{tab:evaluation_metrics}, the KNN model significantly outperforms the GALPROP simulation in the highly structured and dynamically complex Inner Disk/GC region (Zone 3), achieving a remarkable $R^2$ of 0.9465 and reducing the MARE to 14.70\%. The data-driven approach provide crucial insights into dense molecular clouds and unresolved source populations that are challenging for symmetric 3D physical gas models \cite{Guo2018, Gaggero2015}. In contrast, the GALPROP results show slight superiority in the more homogeneous Outer Disk (Zone 4) and South Halo (Zone 2), validating the robustness of the simulation with simpler conditions.

\begin{figure*}[htbp]
    \centering
    \includegraphics[width=0.48\textwidth]{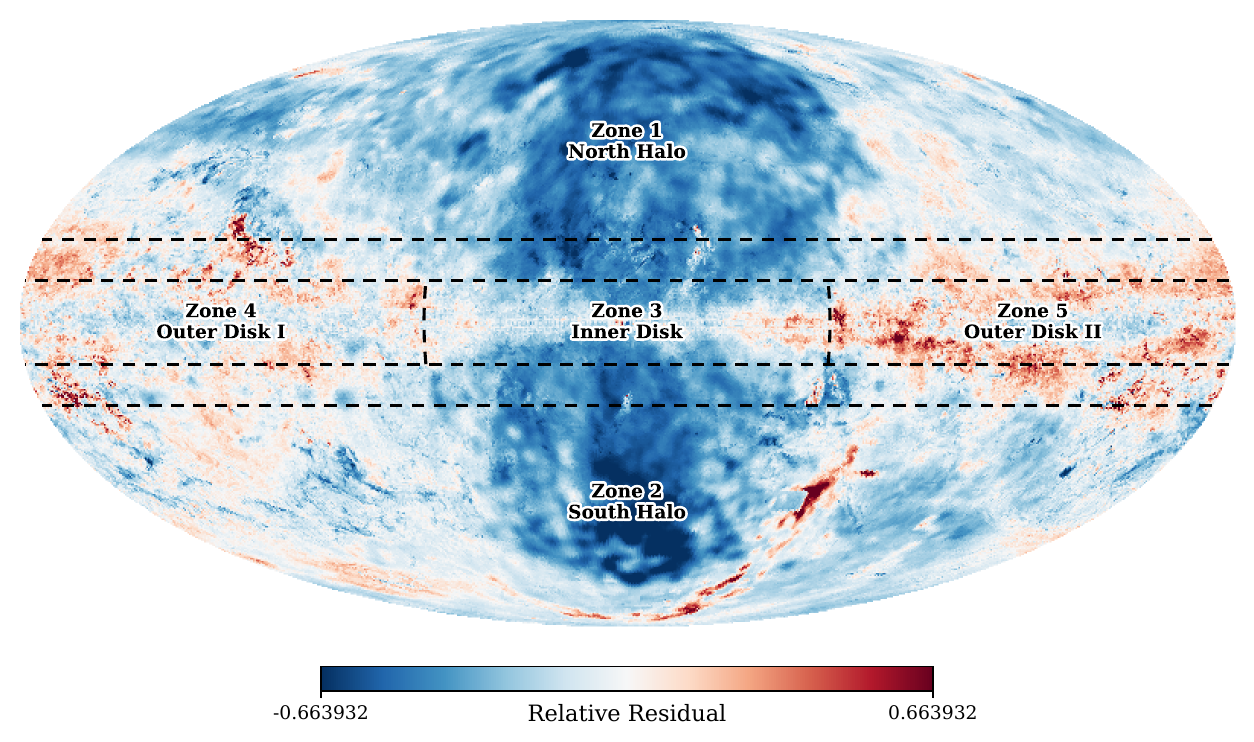}
    \hfill
    \includegraphics[width=0.48\textwidth]{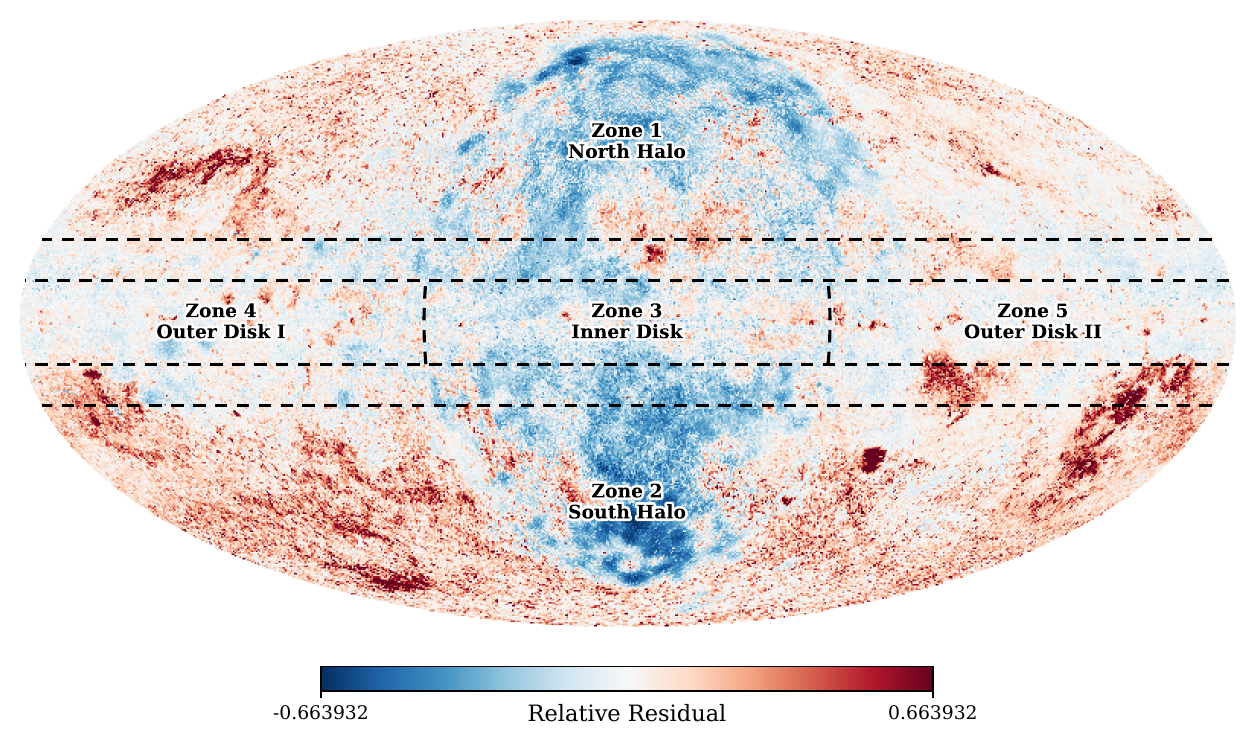}
    \caption{
        Relative residual maps of Galactic gamma-ray emission at $\sim$4.3 GeV. The left panel shows deviations between the GALPROP physical model and \textit{Fermi} observations. Right panel shows deviations between the KNN ML predictions and \textit{Fermi} observations.
    }
    \label{fig:residual_comparison_4300mev}
\end{figure*}

As shown in Fig. \ref{fig:residual_comparison_4300mev}, the GALPROP results reveal evident systematic large-scale deficits along the Galactic plane and near the Inner Galaxy, highlighting the inherent limitations of static gas distributions. In contrast, the ML approach flattens the residuals across the central plane, absorbing non-linear multi-frequency foreground correlations. However, the KNN algorithm is highly sensitive to noise in local extreme pixels, leaving point-like artifacts. In general, this benchmarking confirms the robustness of ML techniques as an indispensable complement to traditional methods.

\section{Discussion and conclusion} \label{sec:conclusion}
In this work, we develop a comprehensive data-driven ML framework to study the connection between Galactic diffuse emission traced by Planck multi-frequency radio/microwave data (30–857 GHz) and \textit{Fermi} gamma-ray observations covering 50 MeV to 814 GeV. By applying RF and KNN regression algorithms, we reconstruct the non-linear mapping between ISM tracers and high-energy emission across the full sky. 

The ML methods achieve excellent performance, with $R^2 > 0.95$ in the 0.1--10 GeV energy range and a maximum $R^2 = 0.96$ at 687 MeV, confirming a tight physical coupling between gas tracers and hadronic gamma-ray emission, especially in the Galactic plane. Cross-hemisphere validation shows that performance varies by less than 6\% across different spatial extents, proving that the radio–gamma relation reflects robust large-scale properties of the Galaxy rather than localized structures. A mild but persistent hemispheric asymmetry appears in the inner Galaxy ($|l|<30^\circ$), likely linked to the Galactic bar and spiral-arm tangents, while the outer disk exhibits high uniformity.

Frequency-dependent analysis reveals that the high-frequency radio bands, dominated by thermal dust emission, provide predictive power nearly identical to that of using all nine frequency bands in the 0.1--10 GeV range. In contrast, low-frequency synchrotron dominated channels perform significantly worse in the low-energy gamma ray, but increasingly better in the high-energy gamma ray. This result offers direct empirical support for the hadronic origin of GeV gamma-ray GDE, where cosmic-ray protons interact with interstellar gas traced by dust, and electron-related processes play a dominant role in the high-energy range for tracing IC process.

Spatial segmentation analysis shows that predictive performance is governed by intrinsic flux similarity rather than spatial proximity. Regions with comparable emission conditions yield accurate cross-prediction even at large distances, while nearby regions with distinct flux levels show poor generalization. The GC represents a special case, where strong structural complexity leads to elevated prediction errors despite high flux similarity between adjacent regions.

Residual maps derived from our radio-based predictions highlight coherent large-scale structures including Loop I, Fermi Bubbles, and the Magellanic Clouds. Positive residuals in the LMC/SMC arise from mismatches between Planck-traced matter distributions and the \textit{Fermi} diffuse model. Negative residuals in Loop I and the NPS indicate additional IC-dominated or locally enhanced high-energy components not captured by radio tracers. The positive residual appears in the north bubble indicating that there might be other components and hybrid orgin needs to be explored for this giant structure \cite{Yang:2018bfb}. These residuals provide a data-driven diagnostic for identifying unmodeled emission and systematic biases in conventional templates.

Benchmarking against the GALPROP shows clear complementary strengths. In the complex inner disk and GC (Zone 3), our KNN model achieves $R^2=0.92$ and reduces the MARE to 18.17\%, significantly outperforming GALPROP ($R^2=0.80$, MARE = 29.16\%). GALPROP remains competitive in the outer disk and halo regions, validating its reliability for GDE study.

This work demonstrates that ML is not just a predictive tool, but a physically interpretable approach. The data-driven baseline established here captures the dominant coupling between ISM tracers and high-energy processes, enabling new constraints on cosmic-ray propagation, interstellar medium structure, and multi-wavelength emission mechanisms.

Future extensions will integrate multi-messenger data, including X-rays, neutrinos, and ultra-high-energy gamma rays, to build a more complete empirical model of Galactic non-thermal emission. Such a framework can help disentangle standard and exotic components, advancing our understanding of high-energy astrophysical processes in the Milky Way.

\begin{acknowledgements}
This work is supported by National SKA Program of China (2025SKA0160100), National Science Foundation of China (12473097, 12261141691, 12005313), the China Manned Space Project with No. CMS-CSST-2021 (A02, A03, B01),  Guangdong Basic and Applied Basic Research Foundation (2024A1515012309).
\end{acknowledgements}

\bibliographystyle{unsrtnat}
\bibliography{references}

\end{document}